\documentclass[aps,prl,twocolumn,superscriptaddress,showpacs]{revtex4}
\usepackage{mathbbold}

\usepackage{graphicx}
\usepackage{mathbbold}
\usepackage{mathbbold}
\begin{document}
\title{\textbf{Generating sub-TeV quasi-monoenergetic proton beam by an ultra-relativistically intense laser in the snowplow regime }}
\author{F.L.Zheng}
\affiliation{Center for Applied Physics and Technology, State Key
Laboratory of Nuclear Physics and Technology,Peking University,
Beijing 100871, China} \affiliation{Graduate School, China Academy
of Engineering Physics, P.O. Box 8009, Beijing 100088, People¡¯s
Republic of China}
\author{H.Y.Wang}
\affiliation{Center for Applied Physics and Technology, State Key
Laboratory of Nuclear Physics and Technology,Peking University,
Beijing 100871, China}
\author{X.Q.Yan}
\email[]{X.Yan@pku.edu.cn} \affiliation{Center for Applied Physics
and Technology, State Key Laboratory of Nuclear Physics and
Technology,Peking University, Beijing 100871, China}
\affiliation{Institute of Applied Physics and Computational
Mathematics, P. O. Box 8009, Beijing 100088, China}
\author{J.E.Chen}
\affiliation{Center for Applied Physics and Technology, State Key
Laboratory of Nuclear Physics and Technology,Peking University,
Beijing 100871, China}
\author{Y.R.Lu}
\affiliation{Center for Applied Physics and Technology, State Key
Laboratory of Nuclear Physics and Technology,Peking University,
Beijing 100871, China}
\author{Z.Y.Guo}
\affiliation{Center for Applied Physics and Technology, State Key
Laboratory of Nuclear Physics and Technology,Peking University,
Beijing 100871, China}
\author{T.Tajima}
\affiliation{Fakult\"at f.\ Physik, LMU M\"unchen, D-85748 Garching,
Germany}
\author{X.T.He}
\email[]{xthe@iapcm.ac.cn}\affiliation{Center for Applied Physics
and Technology, State Key Laboratory of Nuclear Physics and
Technology,Peking University, Beijing 100871, China}

\date{\today}

\begin{abstract}
 Snowplow ion acceleration is presented, using an ultra-relativistically intense laser pulse
irradiating on a combination target, where the relativistic proton
beam generated by radiation pressure acceleration can be trapped
and accelerated by the laser plasma wakefield. The theory suggests
that sub-TeV quasi-monoenergetic proton bunches can be generated
by a centimeter-scale laser wakefield accelerator, driven by a
circularly polarized (CP) laser pulse with the peak intensity of
$10^{23}$W/cm$^2$ and duration of 116fs.

 \end{abstract}
 \maketitle

Laser-driven ion acceleration has drawn attention for many
applications, e.g., proton cancer therapy\cite{therapy}, fast
ignition of thermonuclear fusion by protons
\cite{ignition},conversion of radioactive waste \cite{waster}, high
energy physics accelerator \cite{henp}and astrophysics
\cite{astrophysics}. All these applications may be enabled by the
introduction of near-future high-intensity lasers (i.e. ELI), which
will be capable of producing pulses with intensity $10^{22}-10^{24}$
W/cm$^2$ \cite{ELI}.

 Radiation Pressure Acceleration
(RPA)\cite{rpa} accelerates ions efficiently and theoretical studies
show that GeV proton beam may be generated by an ultra-intense laser
with intensities above $10^{22}$W/cm$^2$
\cite{Esirkepov04,yan09,qiao}. Although the acceleration field is
over tens TeV/m, the acceleration length is normally shorter than
hundreds microns. Because of the energy scaling, it is difficult to
further increase the ion to very high energies (i.e. TeV ).
Quasi-monoenergetic GeV electron beam can be generated by a
centimeter-scale laser plasma wakefield accelerator(LPWA)
\cite{gevelectron}; however, it is not easy to capture and
continuously accelerate the nonrelativistic ions. Shen et
al.\cite{shen09} reports a laser pulse with an ultrarelativistic
intensity can excite an electrostatic field, which can capture
protons from underdense plasma and accelerate them to tens GeV.
Recently Yu et al. \cite{yu10} found that the underdense gas behind
a thin foil can accelerate the proton beam to 60GeV by driving
high-amplitude electrostatic fields moving at a high speed. There is
an important question whether it is possible to generate TeV-level
proton beams by the laser plasma wakefield accelerator.

\begin{figure}
\begin{center}
\includegraphics[width=0.99\columnwidth]{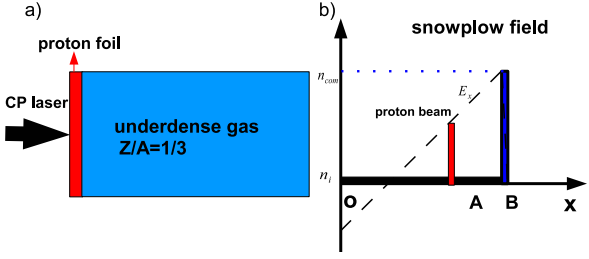}
 \caption{(color online)(a)Acceleration scheme, the initial density of the hydrogen foil $n_0/n_{c}=20$, thickness $D=0.5\lambda_l$;
 the plasma density of the tenuous gas $n_e/n_{c}=0.01$, the length of the gas plasma is $12000\lambda_l$;
(b)Sketch map for the snowplow process, the dynamical density of
ions is $n_i$ and electron density in the snowplow layer is
$n_{com}$. The position $B$ indicates the laser pulse front and $A$
is an arbitrary point for calculating the local electrostatic field
in the snowplow region.}\label{fig.1}
\end{center}
\end{figure}

\begin{figure}
\begin{tabular}{cc}
\includegraphics[totalheight=0.45\columnwidth,width=0.45\columnwidth]{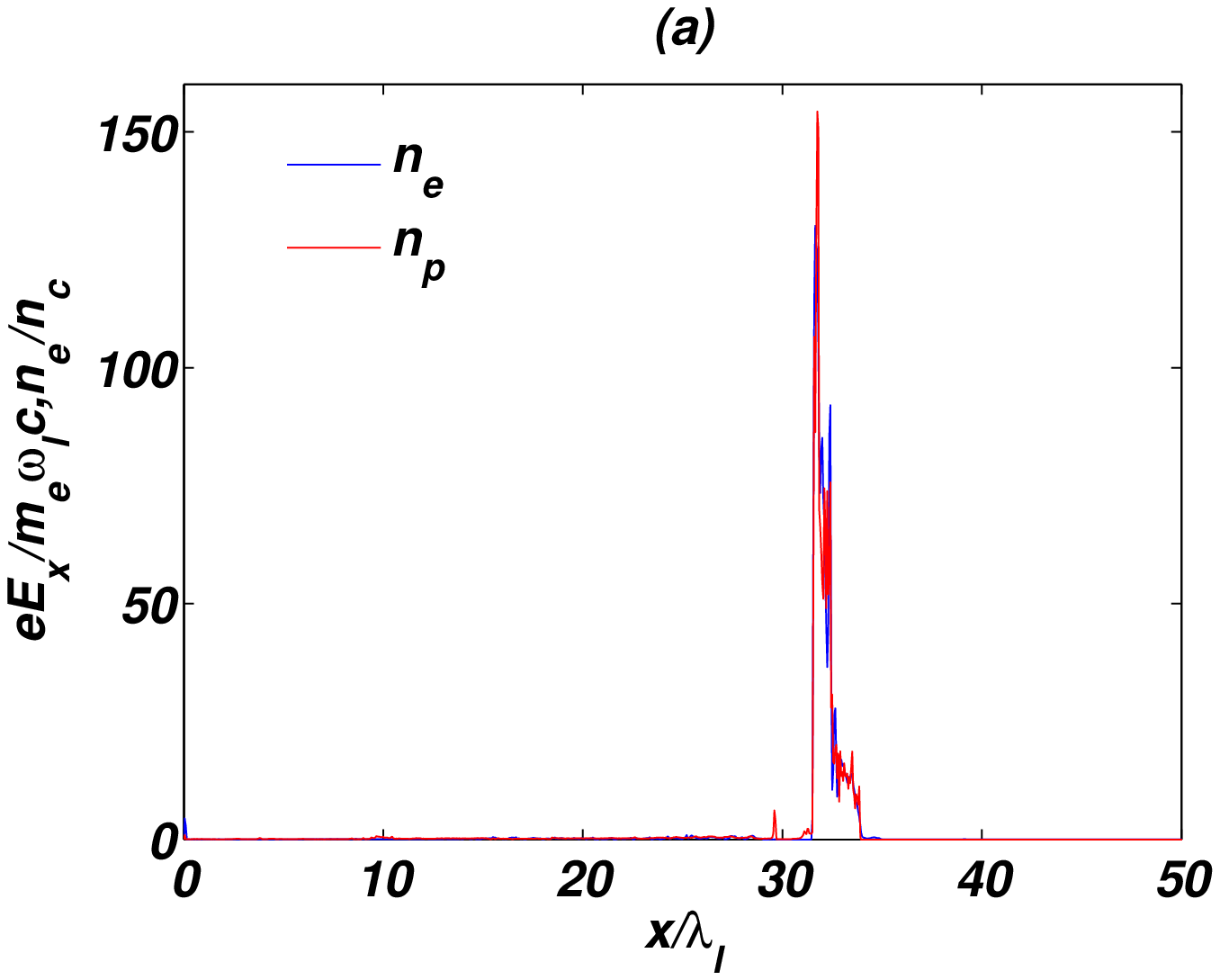}
   &
\includegraphics[totalheight=0.45\columnwidth,width=0.45\columnwidth]{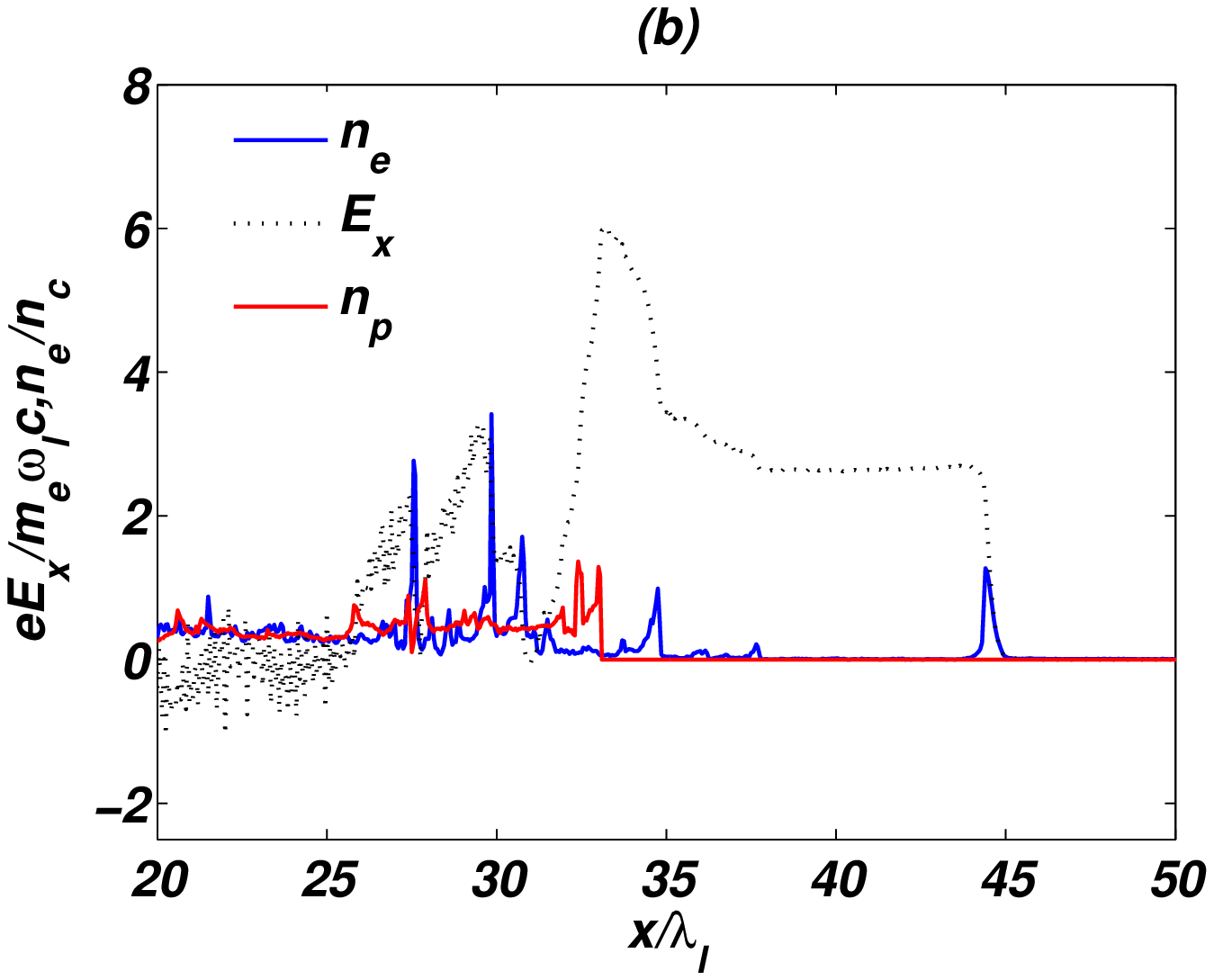}
 \\
\includegraphics[totalheight=0.45\columnwidth,width=0.45\columnwidth]{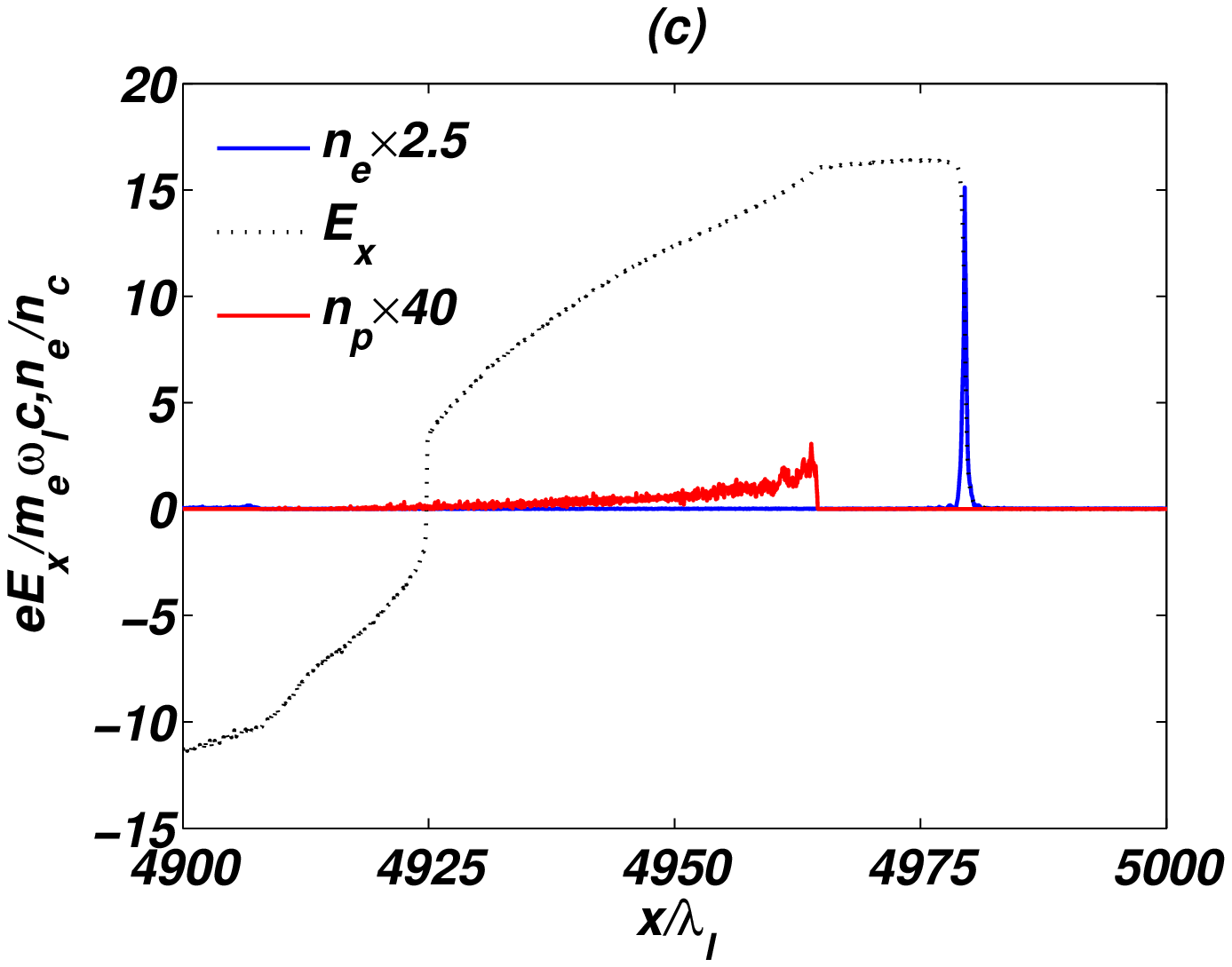}
   &
\includegraphics[totalheight=0.45\columnwidth,width=0.45\columnwidth]{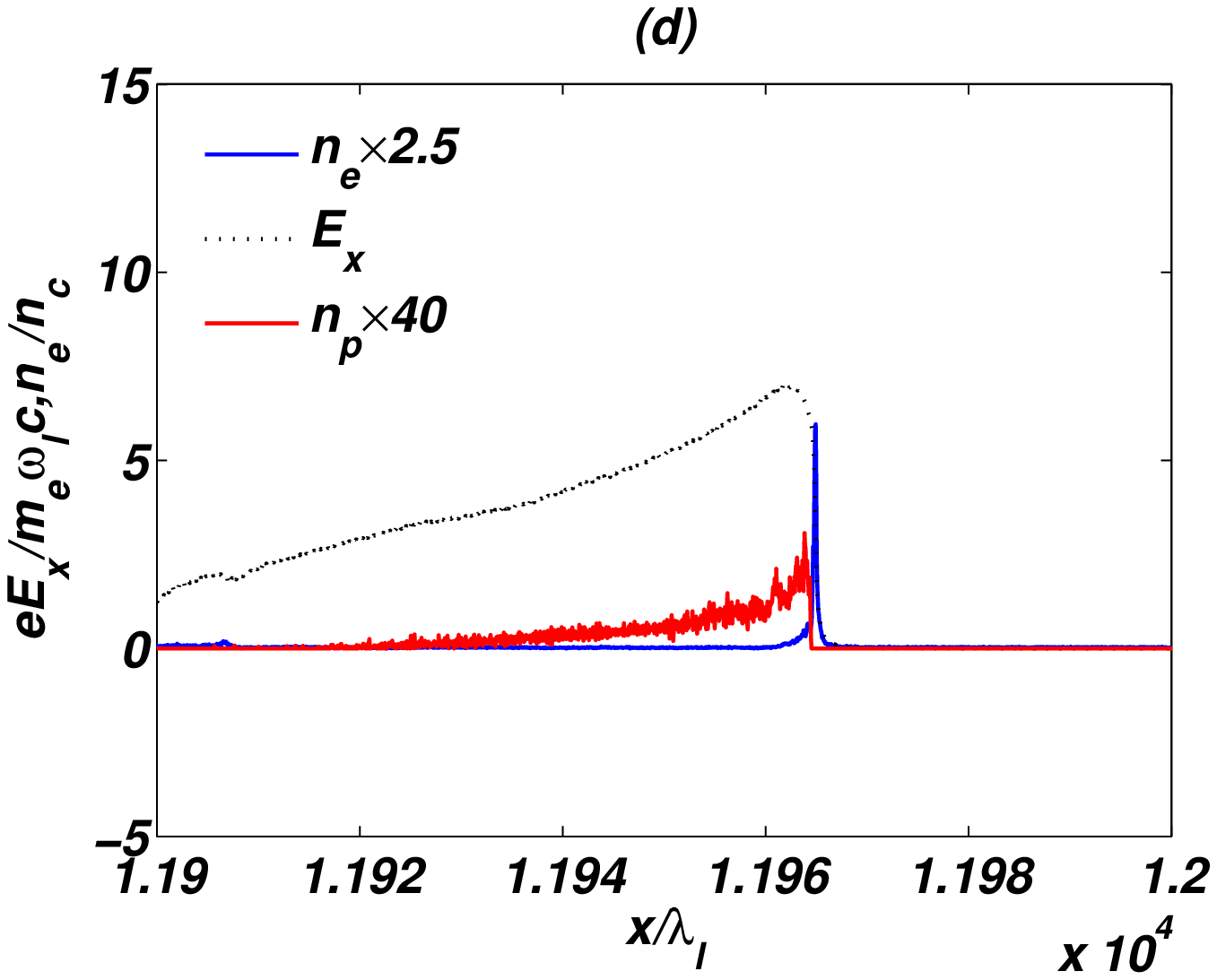}

\end{tabular}
\caption{(Color online)\label{fig.2} Electron density $n_e/n_c$(blue
solid line),proton density $n_p/n_c$(the red solid line) and
longitudinal electrostatic field $eE_x/m_e\omega_lc$ (the black
dashed line).(a)When  $D={1\over 2\pi} {n_c\over n_e}a_0\lambda_l$
double-layer is formed at $t$=50T$_l$. When D=0.5$\mu$m snapshots
are taken at time (b) 50T$_l$, (c) 5000T$_l$, (d)12000T$_l$. }

\end{figure}

This letter reports an ultra-relativistically intense laser can
excite the plasma wakefield that traps the relativistic proton
beam and accelerates it over a few centimeters. An analytic model
is derived for proton acceleration that extends the applicability
of Esarey's classic plasma wakefield theory \cite{Esarey} and has
been confirmed by our PIC simulations. Our scaling law shows that
sub-TeV quasi-monoenergetic proton bunches can be generated by a
circularly polarized (CP) laser pulse with the intensity of
$10^{23}$ W/cm$^2$ and duration of 116 fs. A combined target in
Fig.\ref{fig.1} (a) shows a planar hydrogen target that is
followed by underdense heavy-ion background gas with a
charge-to-mass ratio of $1/3$. When the foil thickness $D$ is
equal to ${1\over 2\pi} {n_c\over n_e}a_0\lambda_l$, a
mono-energetic proton beam is generated from a moving double layer
\cite{yu10}, Fig.\ref{fig.2}(a) suggests that protons and electron
layer are very close. The relativistic protons can quickly overrun
the laser front that results in shorter dephasing length and lower
proton energy. If a thinner foil satisfies
\begin{equation}\label{10} {l_0}< D < {1\over 2\pi}
{n_c\over n_e}a_0\lambda_l,
\end{equation}
 the ponderomotive force quickly pushes
electrons to the rear of the foil (see Fig.\ref{fig.2}(b)), the
electrons that are piled up there reaches density far greater than
$\gamma n_c$. Meanwhile protons in the foil are accelerated by the
separated charge field. Here $l_0$,$n_e$,$n_c$, and $\lambda_l$
are the plasma skin depth, plasma density,critical plasma density,
and laser wavelength, respectively. $a_0=eA/m_e\omega_lc$ is the
normalized laser amplitude and $A$ is the laser vector potential.
$\gamma =(1+I\lambda_l^2/1.37\times10^{18})^{1/2}$, $I$ is laser
intensity, $\lambda_l$ in units of micron. The electron-layer in
rear of the foil is pushed out of the foil by the laser pulse soon
before double-layer (consists of electrons and protons) is formed.
The layer runs in the lower density gas and is further pushed by
the laser pulse like light-sail to generate "Snow-plow" effects on
enhancement of electron number. The electrons in gas plasma drew
away from the foil generate a large electric field, then the
protons that have been accelerated to GeV level start long march
in gas by the wakefield acceleration.
This acceleration scheme is named the snowplow regime. We also
notice a connection of the above plasma acceleration with "snowplow"
acceleration in Ref. \cite{toshi}.
In order to grasp the main physics and to construct a physical model
, in this letter one-dimensional(1D) theory is derived and testified
by 1D and 2D PIC simulations.

Considering the classic 1D plasma wakefield driven by a laser
pulse \cite{Esarey}, two typical features should be noticed in the
ultra-relativistic intensity : (i)Lengthening of the plasma
wavelength $l_{s}=({\gamma n_c/n_e})^{1/2}\lambda_l$;
(ii)Enhancement of the nonlinear wavebreaking field
$E_{B}=3\pi(\gamma n_e/n_c)^{1/2}E_0$, where the critical plasma
density $n_{c}=\pi m_ec^2/e^2\lambda_l^2$. The highly nonlinear
wakefield wavelength $l_s$ and wavebreaking field $E_{B}$ are
increased by a factor of $\gamma^{1/2}$ over the non-relativistic
version, where the electric field strength is normalized by
$E_0=m_e\omega_lc/e$. In this regime we adopt a CP laser pulse
with a wavelength $\lambda_l=1\mu m$ and a peak intensity
$I_0=1.7125\times10^{23}W/cm^2$, corresponding to peak
dimensionless laser amplitude $a_0=250$.
When the foil thickness $D$ satisfies Eq.(1), the relativistic
protons can be captured by the excited wakefield in the underdense
gas and be stably accelerated in the snowplow regime over a long
distance. An analytic model is developed to estimate the
acceleration field, dephasing length, pump depletion length, and
maximum proton energy. The acceleration length in snowplow regime
is proportional to $a_0^{3/2}$ and maximum proton energy scales
with $a_0^2$. It shows that sub TeV monoenergetic proton beam is
generated in the laser intensity of $10^{23}$ W/cm$^2$. To our
best knowledge,it is the best efficient proton acceleration
mechanism in terms of acceleration length and maximum energy.

 First we carried out simulation using a fully relativistic particle-in-cell code (KLAP) \cite{yan09,klap}.
  The laser pulse has a
trapezoidal shape longitudinally with 15 $\lambda_l$ flat top and
20$\lambda_l$ ramp on the rise side. The ramp profile is
$a=a_0sin^2(\pi t/40)$. At time t=0 it is normally incident from
the left on a uniform, fully ionized hydrogen foil of the
thickness $D=0.5\mu m$ and normalized density $N=n_0/n_{c}=20$.
Behind the hydrogen foil is the tenuous heavy-ion-gas with a
charge-to-mass ratio of $1/3$ and a plasma density of
$n_e=0.01n_c$.Here the simulation box has a size of 12000
$\lambda_l$ in x plane and a space resolution of 20
cells/$\lambda_l$. Each cell is filled with 10 macroparticles for
gas plasma and 20000 macroparticles for foil plasma. The initial
electron and ion temperature are 0.1eV.

After the proton beam is accelerated to GeV level in the laser foil
interactions, the laser pulse passes through the foil and piles
electrons in the underdense gas plasma as Fig.\ref{fig.1} (b) shows,
which is one of most important and remarkable features in the
snowplow regime. Here the electron layer and the positive
electrostatic field are called the snowplow layer and snowplow
field, respectively.
The snowplow layer can continuously trap electrons in the snowplow
process, so the acceleration field exists stably for a long
distance. In the snowplow region, it is found that the length of the
snowplow region is equal to $l_s=({\gamma n_c/n_e})^{1/2}\lambda_l$
and the snowplow velocity approximately equals to the group velocity
of laser pulse in the underdense plasma.
First we estimate the electrostatic field $E_A$ at an arbitrary
point $x_A$ in the snowplow region. The component from ions is
$E_A^{'}={4\pi en_ix_{A}}-{4\pi en_i(l_s-x_{A})}$.
The contribution from the snowplow layer at position $x_A$ is
 $E_A^{''}={4\pi en_{e}l_s\over 2}$.
The longitudinal electrostatic field at $x_A$ is
$E_A=E_A^{'}+E_A^{''}={{4\pi en_{e}}(2x_{A}-l_s/2)}$. Therefore, the
maximum snowplow field at the laser front $x_B$ is
\begin{equation}\label{4}
E_{B}=3\pi(\gamma n_e/n_c)^{1/2}E_0.
\end{equation}

Since the injected GeV protons are still slower than the laser
pulse at the beginning, the ion beam should be injected into the
front side of snowplow region. Afterwards protons are gradually
accelerated and overrun the laser pulse ( see Fig.\ref{fig.4}(a)).
The optimized distance between injected proton beam and laser
pulse front is defined by $l_{inj}$. It is approximately
${1\over4}l_s$ in our simulation. Once the proton beam is injected
into the wakefield, its maximum energy depends on the laser pump
depletion length and proton dephasing length.
The laser pump depletion length $L_{pd}$ is estimated
\cite{Esarey}:
$E_x^2 \cdot L_{pd}\simeq E_l^2\cdot L_{l}$,
where $E_x$ is the longitudinal electrostatic field driven by laser
pulse in the snowplow wakefield and $L_l$ is the length of laser
pulse. Assuming the group velocity of laser pulse is extremely close
to light velocity in free space, pump depletion length can be
defined as
\begin{equation}\label{lpd}
L_{pd}\simeq {L_{l}\over v_{etch}}\cdot c,
\end{equation}
where $v_{etch}$ is the depletion velocity of laser pulse. The
etching velocity of laser pulse reads
$v_{etch}\simeq \frac {9\pi^2 n_e}{a_0n_c}  c$.
It shows the etching velocity of laser pulse scales as $n_e/a_0$.

The dephasing length of proton is given approximately by
$L_{dp}\simeq{l_{inj }\over v_{p}-(v_{g}-v_{etch})}\cdot v_{p}$,
where $l_{inj}$ is the injected length between proton beam and
laser pulse front, $v_{g}$ is the laser group velocity, $v_{p}$ is
the proton velocity. It can be further simplified when both
$v_{g}$ and $v_{p}$ are very close to light speed, which means
$v_{etch}\gg (v_{p}-v_{g})$. The proton dephasing length is:
\begin{equation}\label{Ldp}
L_{dp} \simeq {\frac {1}{36\pi^2} (a_0 n_c/n_e)^{3/2} \lambda_l}.
\end{equation}
In case of $L_l > \frac {1}{4}(a_0n_c/n_e)^{1/2}\lambda_l$ the laser
pump depletion length is greater than the dephasing length , the
maximum energy proton beam can be given as
\begin{equation}\label{Wmax}
W_{max} \simeq \frac {1}{6} (a_0^2n_c/n_e)m_ec^2.
\end{equation}
It implies that the maximum proton energy in the snowplow regime
scales with $a_0^{2}$ and $n_e^{-1}$. This is quite an efficient
acceleration scaling law.
\par
\begin{figure}
\begin{tabular}{cc}
\includegraphics[totalheight=0.45\columnwidth,width=0.45\columnwidth]{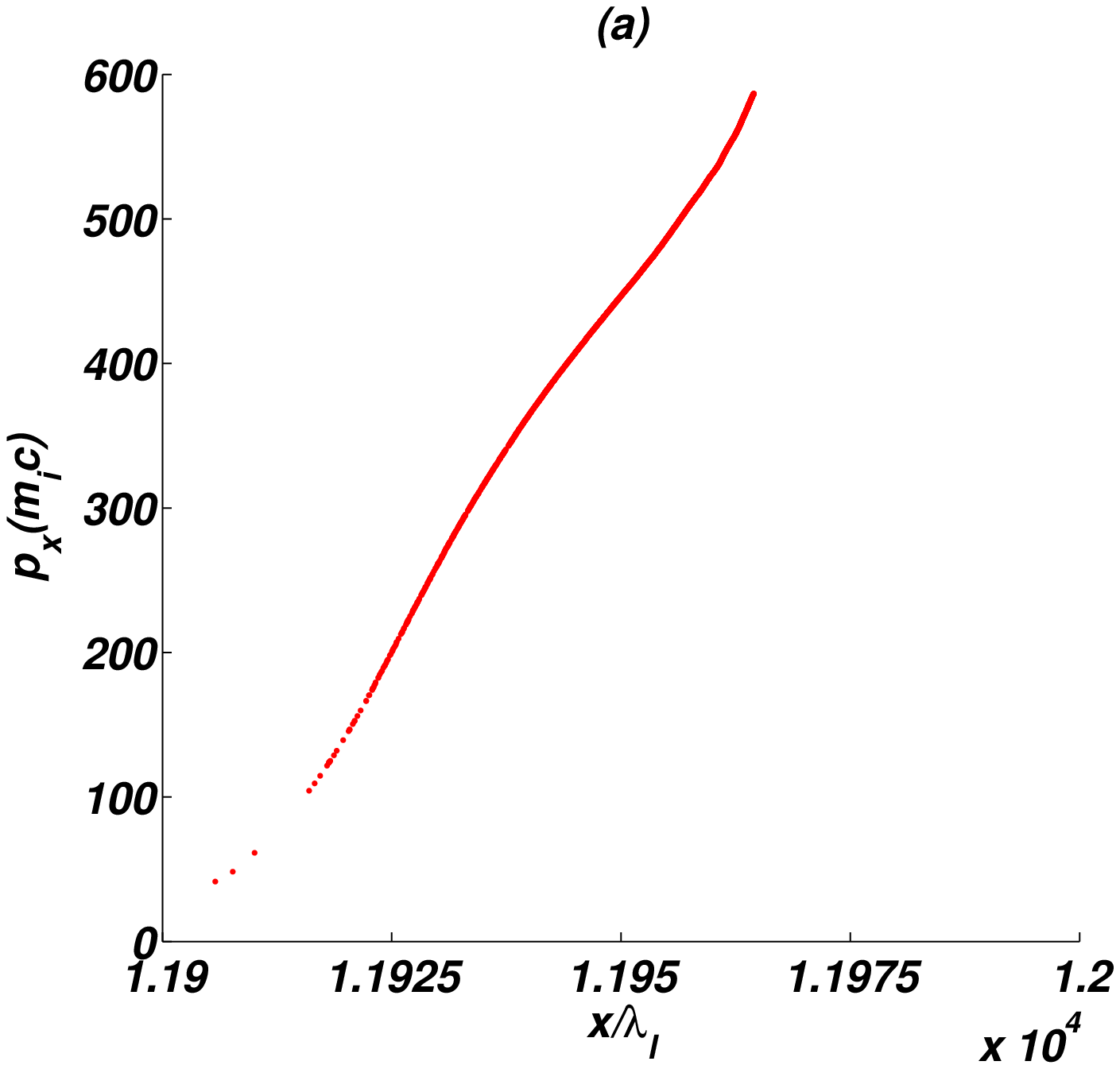}
   &
\includegraphics[totalheight=0.45\columnwidth,width=0.45\columnwidth]{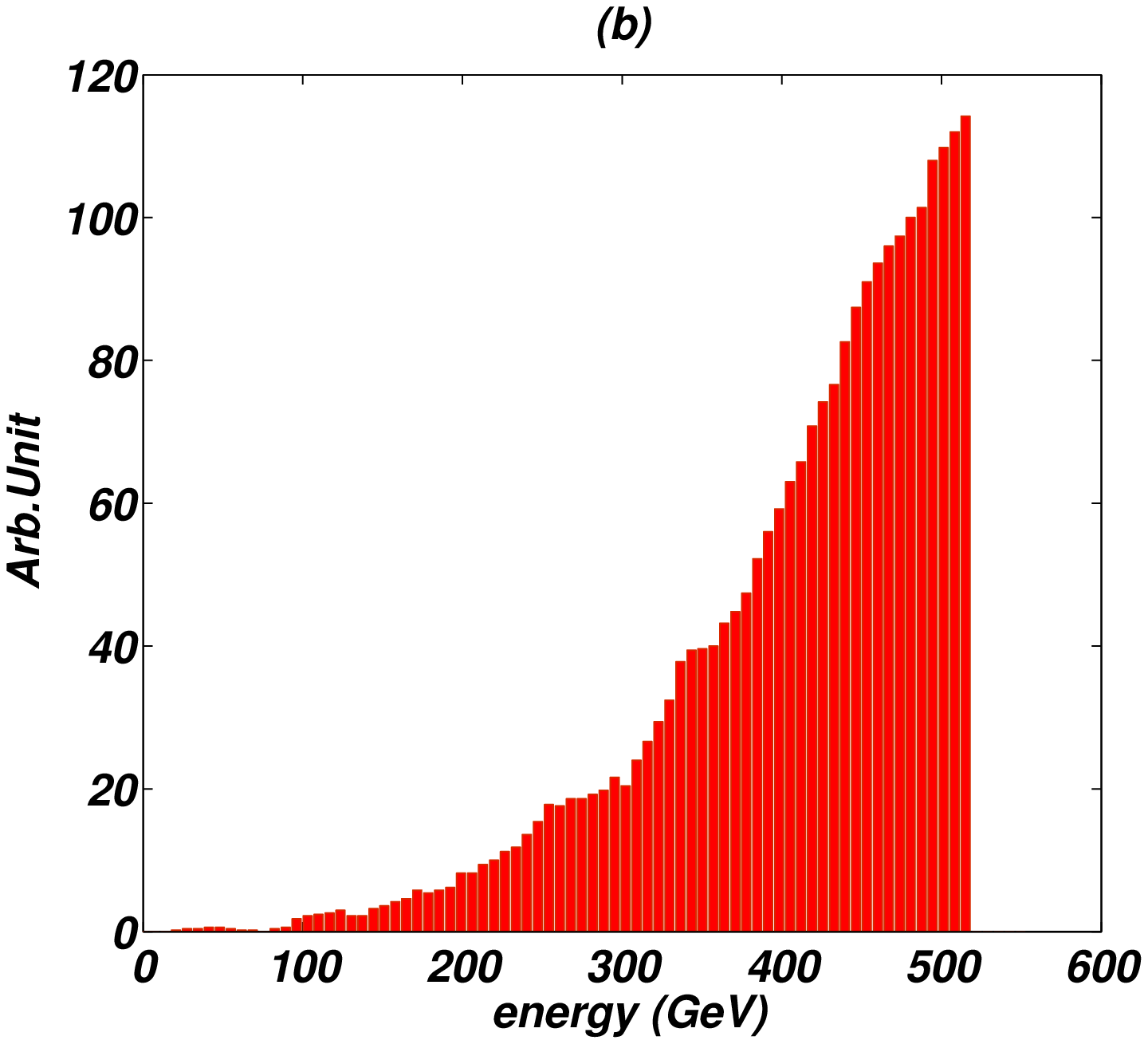}
 \\
\end{tabular}
\caption{(Color Online) Longitudinal phase space for protons
$p_x/m_ic$ (a) and energy spectrum of trapped protons (b) from
simulations at $t$=12000T$_l$.}\label{fig.3}
\end{figure}
\begin{figure}
\begin{center}
\includegraphics[width=0.99\columnwidth]{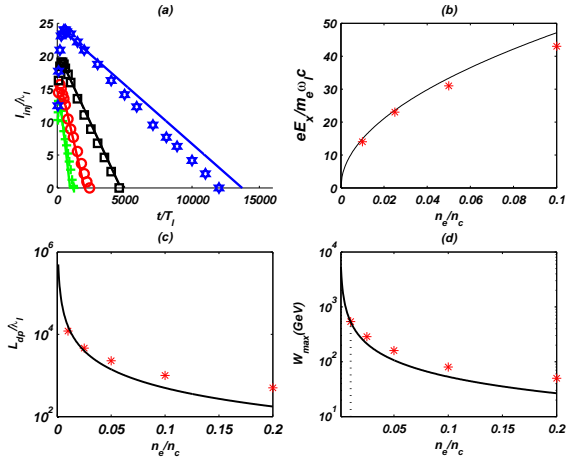}
\caption{(Color Online)(a)Distance between proton beam and the laser
pulse front. It varies with time for different gas densities, where
blue, black, red and green curves represent the gas densities of
$0.01n_c$, $0.025n_c$, $0.05n_c$ and $0.1n_c$, respectively;
(b)Longitudinal electrostatic field $eE_x/m_e\omega_lc$;(c)Dephasing
length; (d) Maximum proton energy. Here the solid lines are
theoretical curves and simulation results are marked by different
symbols(e.g. star,diamond and squre).}\label{fig.4}
\end{center}
\end{figure}

The snapshots of electron density, proton density, and electrostatic
field are plotted in Fig.\ref{fig.2}. The proton beam has been
continuously accelerated until laser pulse is completely depleted at
$t$=12000T$_l$. When the proton beam approaches the laser front, the
beam loading causes a substantial reduction of the electrostatic
field as shown in Fig.\ref{fig.2}(c) and \ref{fig.2}(d). The
theoretical values of the longitudinal electrostatic field
$E_x=14.9$ in the unit of $E_0$, the dephasing length
$l_{pd}=11125\mu m$,the maximum energy of proton beam $W_{max}=532$
GeV and the etching velocity $v_{etch}=0.0036c$,respectively. They
are consistent with the simulation results. Fig.\ref{fig.3} shows
the energy spread (FWHM ) of the beam is less than 20 $\%$. The
maximum proton energy and averaged one are 540GeV and 437GeV,
respectively. There are $1.72\%$ of the total protons located inside
this energy window. The proton beam is compressed in the phase space
and the final quasi-monoenergetic beam is shorter than 100 $\mu m$.

As Fig.\ref{fig.4}(a) shows, proton beams move backward with
respect to the snowplough reference since the laser group velocity
is greater than the proton beam. As a result, the distance between
proton beam and laser front increases to reach the maximum
l$_{inj}$ in the initial injection stage.
Later it begins to decrease due to the erosion
of the laser pulse front.
 The slopes of the solid lines represent the etching
velocities of the laser pulse for different gas densities.
The maximum longitudinal electrostatic field, dephasing length,
and maximum proton energy are plotted versus the density in
Fig.\ref{fig.4}(b,c,d). It shows the maximum energy of the proton
beam is increasing with plasma density decreasing. The dotted
lines indicate that the pulse depletion length is shorter than the
dephasing length if the plasma density is smaller than $0.01n_c$.
In order to increase the proton energy, we may increase the
intensity of lase pulse and reduce the gas density, which can
increase both the dephasing length and pump depletion length.


The density ranges over more than three orders of magnitude in 1D
simulations, it is difficult to do a full scale 2D simulations (e.g.
TeV protons). A small simulation box 40$\times$1000$\lambda_l^2$ is
chosen with a spatial resolution of 5 cells/$\lambda_l$ in y plane
and 20 cells/$\lambda_l$ in x plane.
Each cell contains 4 particles for each species of gas plasma, 400
particles for foil plasma. The foil is the same as in 1D simulations
while the gas density is 0.2$n_c$. The laser pulse is temporally and
transversely super-Gaussian, $I=I_0
exp(-(r/r_{0})^4-[(t-t_{0})/\tau]^8)$, where $r_0$=10 $\mu$m,
$t_0$=55 fs, and $\tau$=55 fs are taken.
 Fig.{\ref{fig.5} shows the snowplow layer,electrostatic field,and phase space
 are similar with 1D results while the maximum proton energy reaches 50GeV that
agrees with 1D simulations. 
Since protons gain energy mainly from the electrostatic field
\cite{yu10}, the return current and magnetic field in 2D simulations
are not important for proton acceleration. Furthermore, the lost
fraction of electrons in snowplow layer can be compensated from gas
plasma and snowplow layer is still kept at the end of simulations.
The disadvantageous multi-dimensional effects (e.g. hole boring,
Rayleigh-Taylor instability) in laser foil interactions \cite{rpa}
are not observed in our simulations. These effects lead to very
stable proton acceleration in the multi-dimensional situation.

\begin{figure}
\begin{tabular}{cc}
\includegraphics[totalheight=0.36\columnwidth,width=0.45\columnwidth]{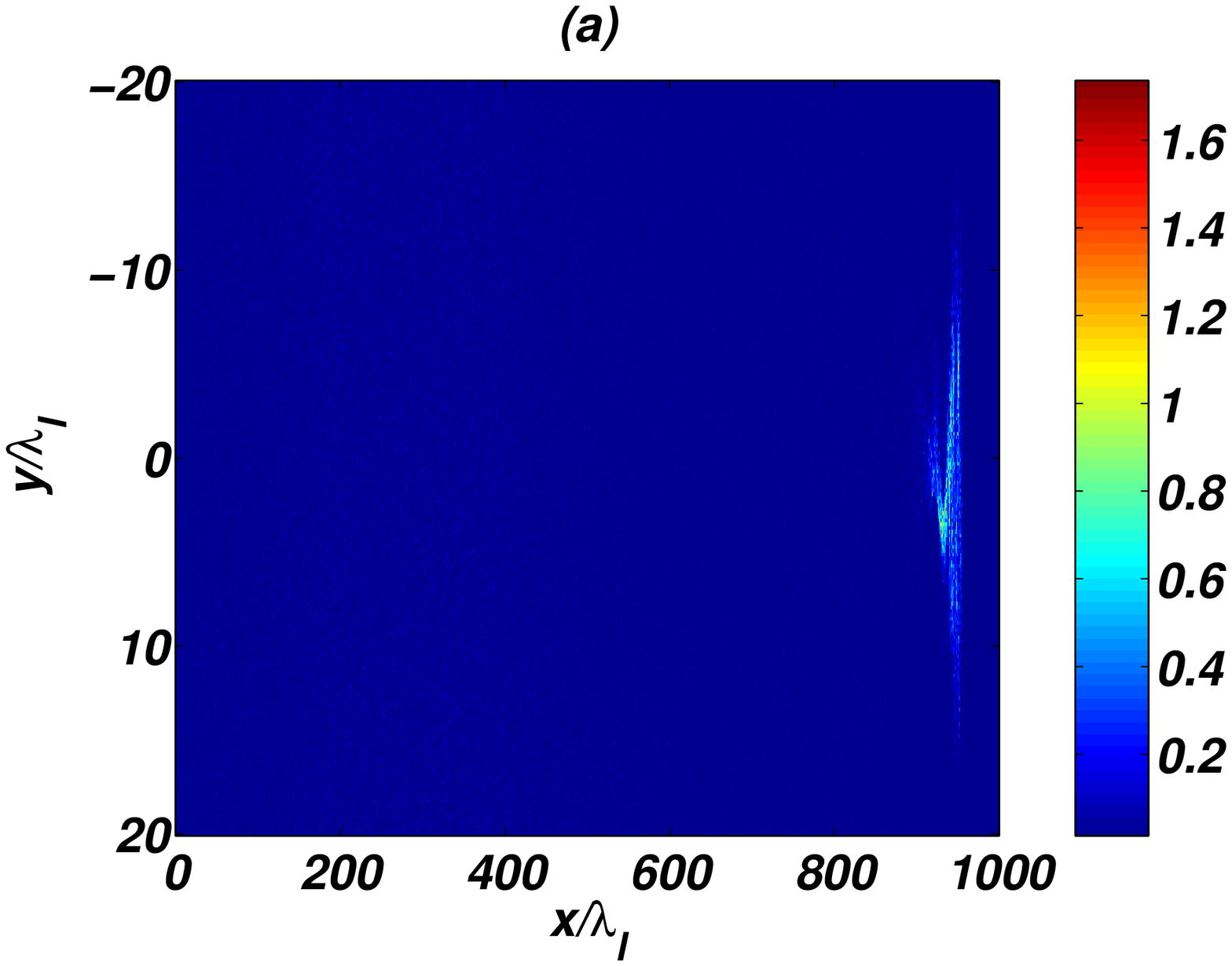}
   &
\includegraphics[totalheight=0.36\columnwidth,width=0.45\columnwidth]{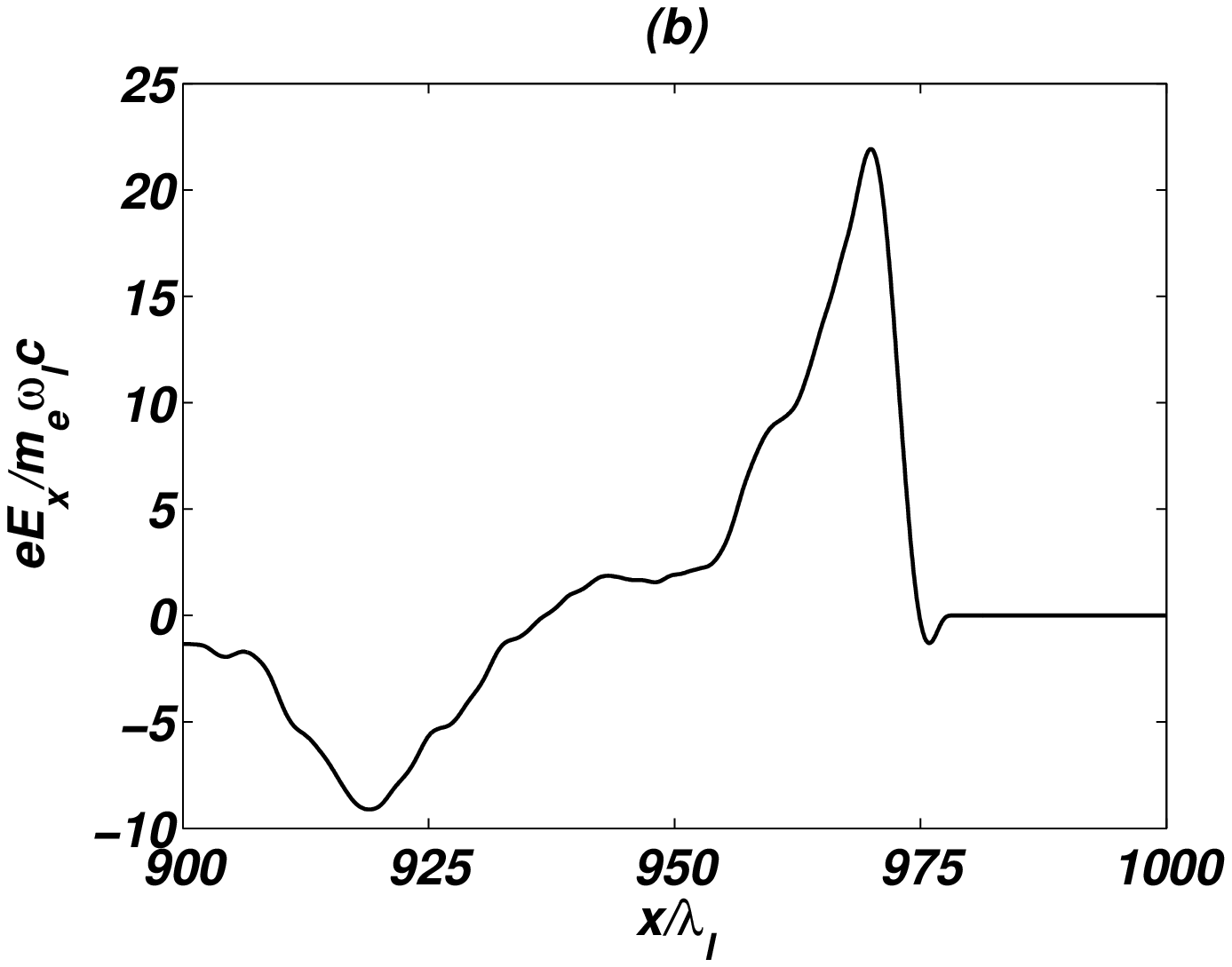}
 \\
\includegraphics[totalheight=0.36\columnwidth,width=0.45\columnwidth]{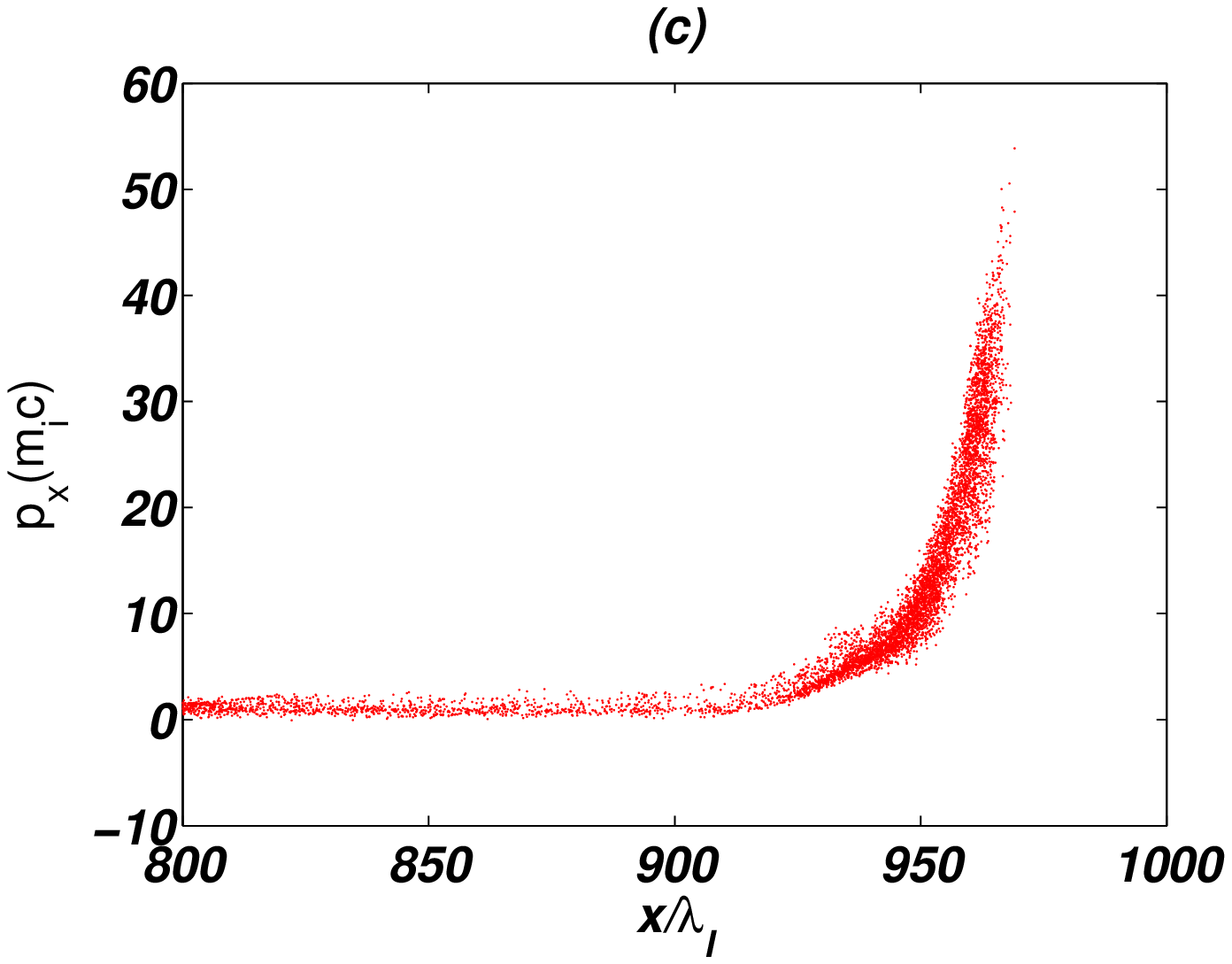}
   &
\includegraphics[totalheight=0.36\columnwidth,width=0.45\columnwidth]{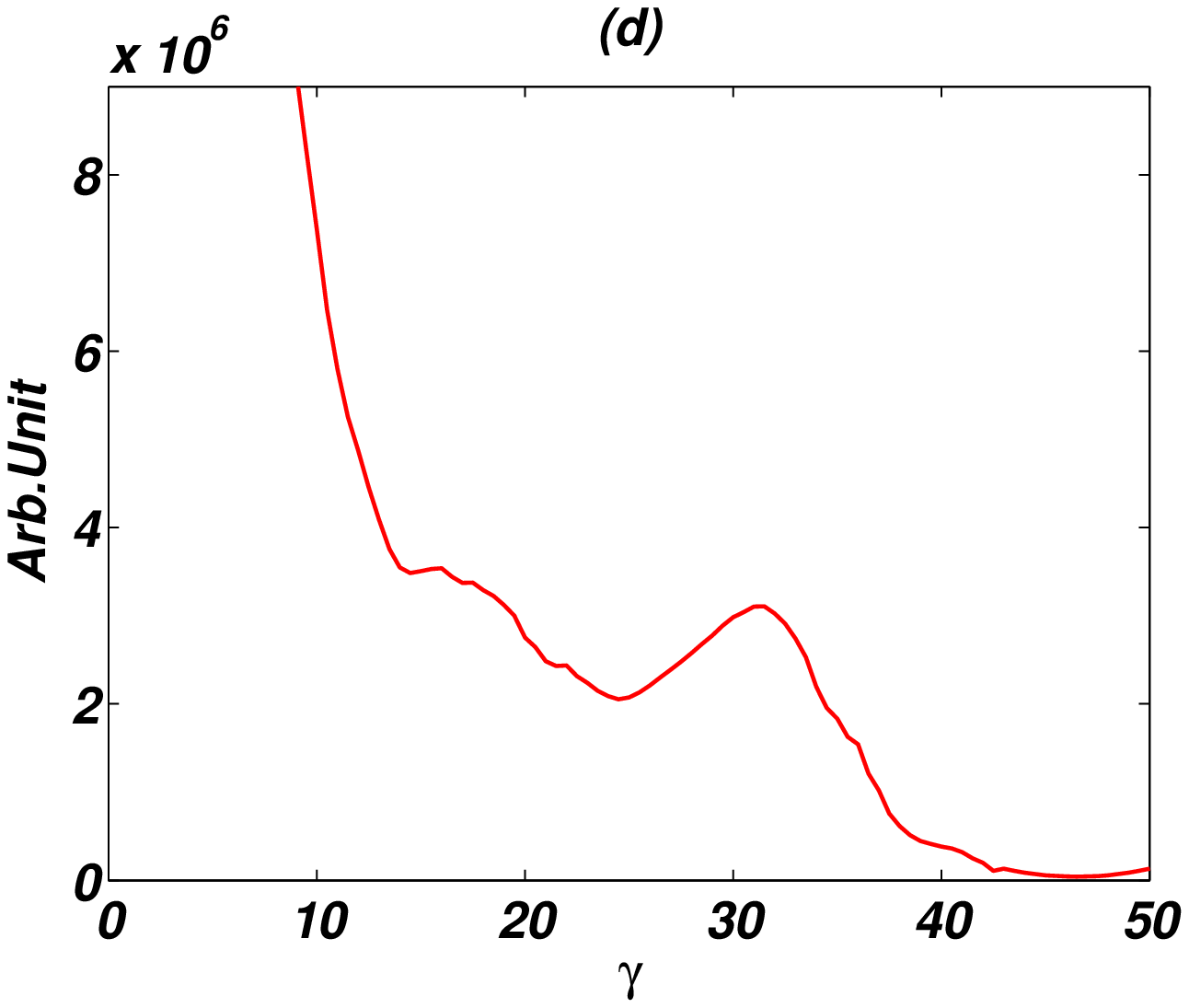}

\end{tabular}
 \caption{(Color Online)2D simulation results at t=990 $T_l$.(a)Electron density
 distribution; (b)Electrostatic field on the axis
 ;(c)Proton phase space; (d)Proton spectrum.}\label{fig.5}
\end{figure}

In conclusion,  snowplow ion acceleration, using an
ultra-relativistically intense laser pulse irradiating on the
combination target, is presented. A 1D model that can predict the
proton dephasing length, pump depletion length, and  maximum proton
energy is derived, which is verified by 1D and 2D PIC simulations.
It shows that sub-TeV quasi-monoenergetic proton bunches can be
generated by a centimeter-scale laser wakefield accelerator, excited
by a laser pulse with the intensity of $10^{23}$W/cm$^2$ and
duration of 116fs. The final proton beam is shorter than 100 $\mu$m
and may be used to excite the plasma wakefield to accelerate
electrons to hundreds GeV, as Cadwell proposed \cite{allen}.


We thank G.Mourou, Z.M.Sheng, C.T.Zhou, C. Y. Zheng, H. Zhang, B.
Liu, and H.C.Wu for useful discussions and help.This work was
supported by National Nature Science Foundation of China (Grant Nos.
10935002,10835003,11025523) and National Basic Research Program of
China (Grant No. 2011CB808104). XQY would like to thank the support
from the Alexander von Humboldt Foundation. TT is the holder of the
Blaise Pascal chair of Ecole Normale Superieure.


\bf
\end{document}